\documentclass{iau}
\usepackage{graphicx}
\usepackage[comma,authoryear]{natbib}
\usepackage[latin1]{inputenc}
\usepackage{hyperref}

\title[Pulsations of rapidly rotating stars with compositional discontinuities]
      {Pulsations of rapidly rotating stars with compositional discontinuities}

\author[D. R. Reese, F. Espinosa Lara \& M. Rieutord]
{Daniel R. Reese$^1$
 \and Francisco Espinosa Lara$^{2,3}$
 \and Michel Rieutord$^{2,3}$}

\affiliation{$^1$Institut d'Astrophysique et G{\'e}ophysique de l'Universit{\'e} de Li{\`e}ge , \\
           All{\'e}e du 6 Ao{\^u}t 17, 4000 Li{\`e}ge, Belgium \\ email: {\tt daniel.reese@ulg.ac;be} \\[\affilskip]
           $^2$Universt{\'e} de Toulouse, UPS-OMP, IRAP, Toulouse, France \\[\affilskip]
           $^3$CNRS, IRAP, 14 avenue Edouard Belin, 31400 Toulouse, France}

\pubyear{2013}
\volume{301}  
\pagerange{119--126}
\jname{Precision Asteroseismology. Celebration of the Scientific Opus of Wojtek Dziembowski}
\editors{W. Chaplin, J. Guzik, G. Handler \& A. Pigulski, eds.}
\begin{document}

\maketitle

\begin{abstract}
Recent observations of rapidly rotating stars have revealed the presence of
regular patterns in their pulsation spectra. This has raised the question as to
their physical origin, and in particular, whether they can be explained by an
asymptotic frequency formula for low-degree acoustic modes, as recently
discovered through numerical calculations and theoretical considerations.  In
this context, a key question is whether compositional/density gradients can
adversely affect such patterns to the point of hindering their identification. 
To answer this question, we calculate frequency spectra using two-dimensional
ESTER stellar models. These models use a multi-domain spectral approach,
allowing us to easily insert a compositional discontinuity while retaining a
high numerical accuracy. We analyse the effects of such discontinuities on both
the frequencies and eigenfunctions of pulsation modes in the asymptotic regime.
We find that although there is more scatter around the asymptotic frequency
formula, the semi-large frequency separation can still be clearly identified in
a spectrum of low-degree acoustic modes.
\end{abstract}

\firstsection 
\section{Introduction}

Recent observations of pulsation spectra in rapidly rotating stars have revealed
the presence of frequency patterns.  For instance, \citet{GarciaHernandez2009,
GarciaHernandez2013} found recurrent frequency spacings in two $\delta$ Scutis
observed by CoRoT, thereby allowing the construction of an \textit{echelle}
diagram in the latter case.  Similarly,  \citet{Breger2012, Breger2013} found
multiple sequences of very uniformly spaced frequencies in a $\delta$ Scuti
observed by Kepler.  These observations show that although the pulsation spectra
of $\delta$ Scuti stars lack the \emph{simple} frequency patterns present in
solar-type pulsators, regular patterns do exist in such stars and need to be
explained.

Among the various possible explanations, one particularly interesting option is
the asymptotic frequency pattern for low-degree acoustic modes (\textit{i.e.}
island modes) in rapidly rotating stars, recently discovered through numerical
\citep{Lignieres2006, Reese2008a, Reese2009a} and theoretical considerations
\citep{Lignieres2008, Lignieres2009, Pasek2011, Pasek2012}.  Identifying such a
pattern in rapidly rotating stars could yield useful information such as the
mean density \citep{Reese2008a, GarciaHernandez2013}.  However, an open question
is up to what extent it is affected by strong gradients or glitches (such as
$\mu$ gradients, ionisation zones, or boundaries of convective regions), and
whether this can hinder its identification.  In order to answer this question,
we investigate the pulsation spectra of rapidly rotating models with sharp
discontinuities.

\section{Numerical calculations}

\begin{figure}[htbp]
\begin{center}
\includegraphics[height=4cm]{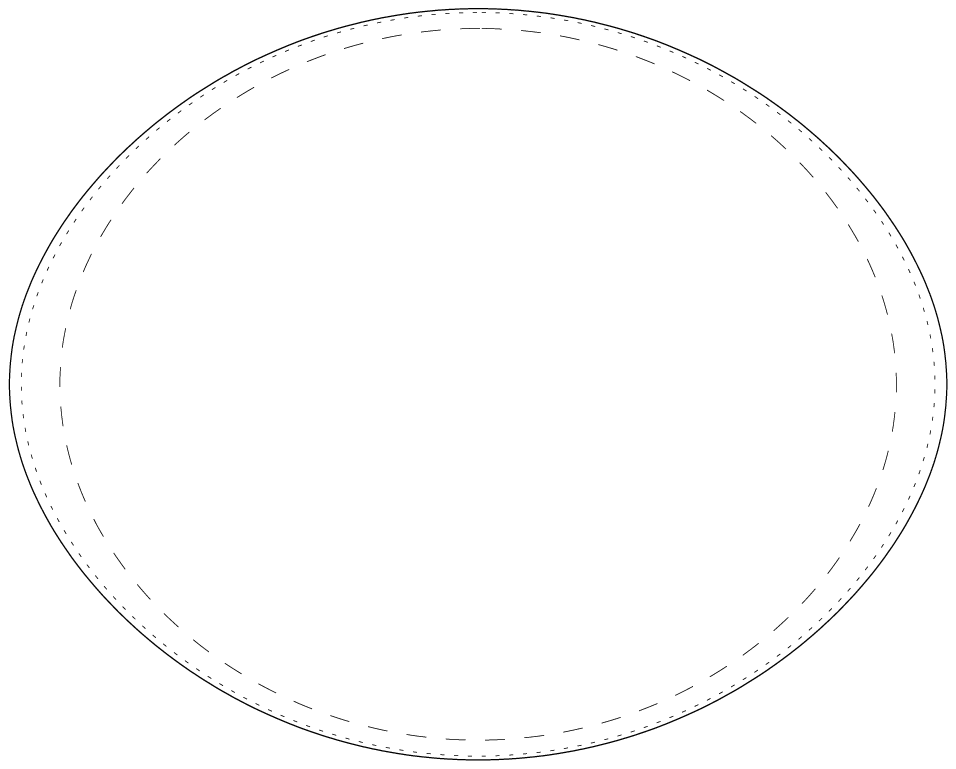} $\quad$
\includegraphics[height=4cm]{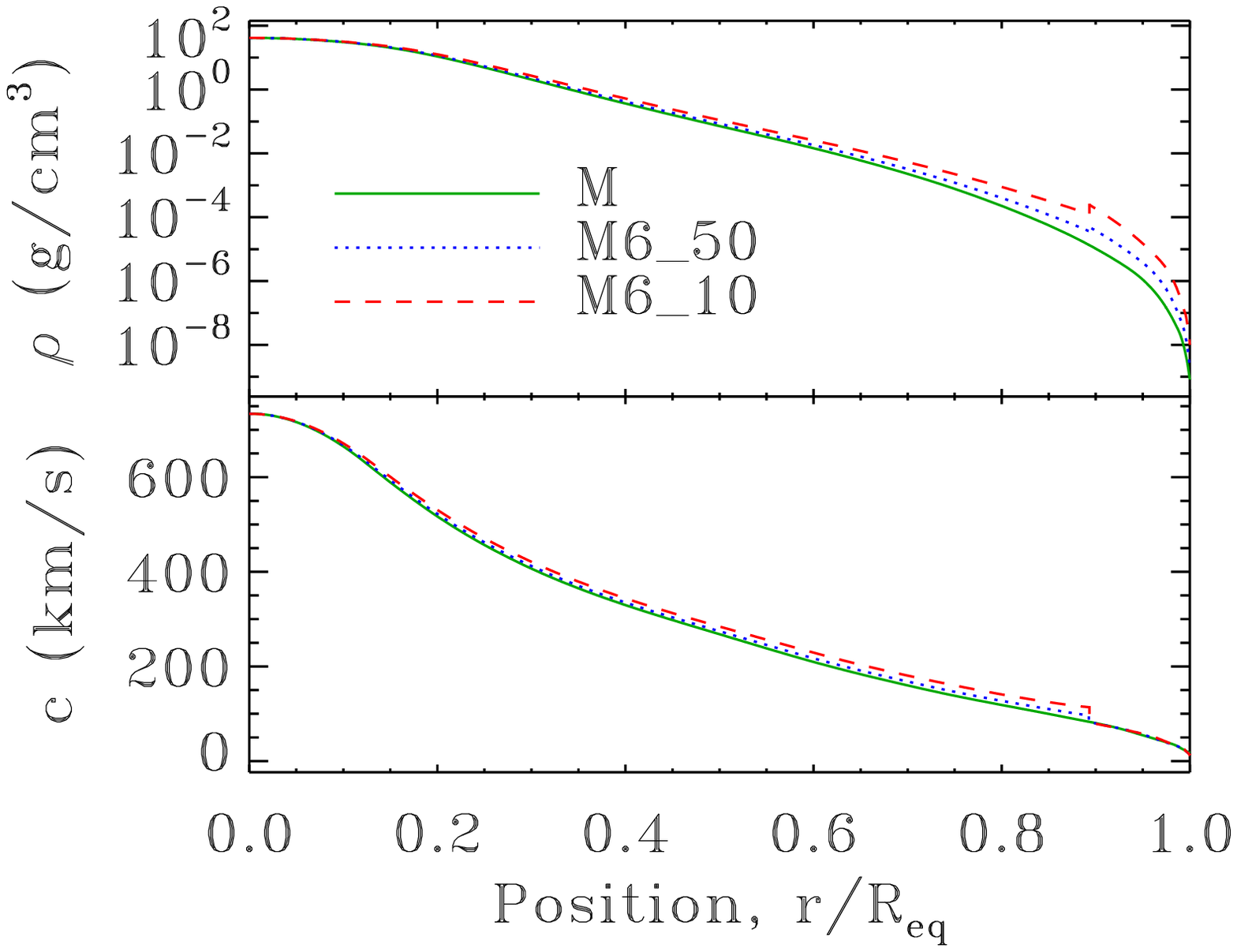}
\end{center}
\caption{\textit{Left:} meridional cross-sections of discontinuities in models
\texttt{M6\_xx} (dashed line) and \texttt{M7\_xx} (dotted line), and stellar
surface (solid line).  \textit{Right:} density and sound velocity profiles for
three of the models. \label{fig:profiles_domains}}
\end{figure}

We worked with various 3 $M_{\odot}$ models, where the surface is rotating at
$70\,\%$ of the Keplerian break-up rotation rate ($v_{\mathrm{eq}} = 340 - 350$
km/s). These models were produced by the 2D multi-domain spectral code ESTER,
which self-consistently calculates the rotation profile, $\Omega$
\citep{Rieutord2009, Rieutord2013, EspinosaLara2013}.  Its multi-domain approach
is well-suited to introducing discontinuities without sacrificing numerical
accuracy, since these can be made to coincide with domain boundaries.  In what
follows, we worked with five different models: \texttt{M} which is smooth,
\texttt{M6\_50}, \texttt{M6\_10}, \texttt{M7\_50}, and \texttt{M7\_10}.  Models
\texttt{M6\_xx} have a discontinuity deeper within the star (see
Fig.~\ref{fig:profiles_domains}, left panel).  In all cases, the discontinuities
follow isobars.  The surface hydrogen content is decreased by $50\%$ and $90\%$
in models \texttt{Md\_50} and \texttt{Md\_10}, and corresponds to a $17\%$ and
$39\%$ jump in the speed of sound, respectively (see
Fig.~\ref{fig:profiles_domains}, right panel).

Adiabatic calculations of acoustic pulsation modes were carried out thanks to
the TOP code which fully takes into account the effects of rotation
\citep{Reese2009a}.   Regularity conditions were applied in the centre, the
simple mechanical condition $\delta p = 0$ was enforced at the stellar surface,
and the perturbation to the gravity potential was made to vanish at infinity. 
Various matching conditions were needed to ensure that the perturbation of the
pressure, the gravity potential, and its gradient, remain continuous across the
\textit{perturbed} discontinuity.  Furthermore, the fluid domain had to be kept
continuous by making sure that the deformation caused by the fluid displacement
is the same below and above the discontinuity. Similar calculations had
previously been carried out in \citet{Reese2011}.  However, these calculations
did not take into account the fact that the matching conditions apply across the
\textit{perturbed} discontinuity, and the results were less conclusive because
the discontinuity was located deeper within the star, where acoustic island
modes are less sensitive.

\section{Results}

\begin{figure}[htbp]
\begin{center}
\includegraphics[height=4cm]{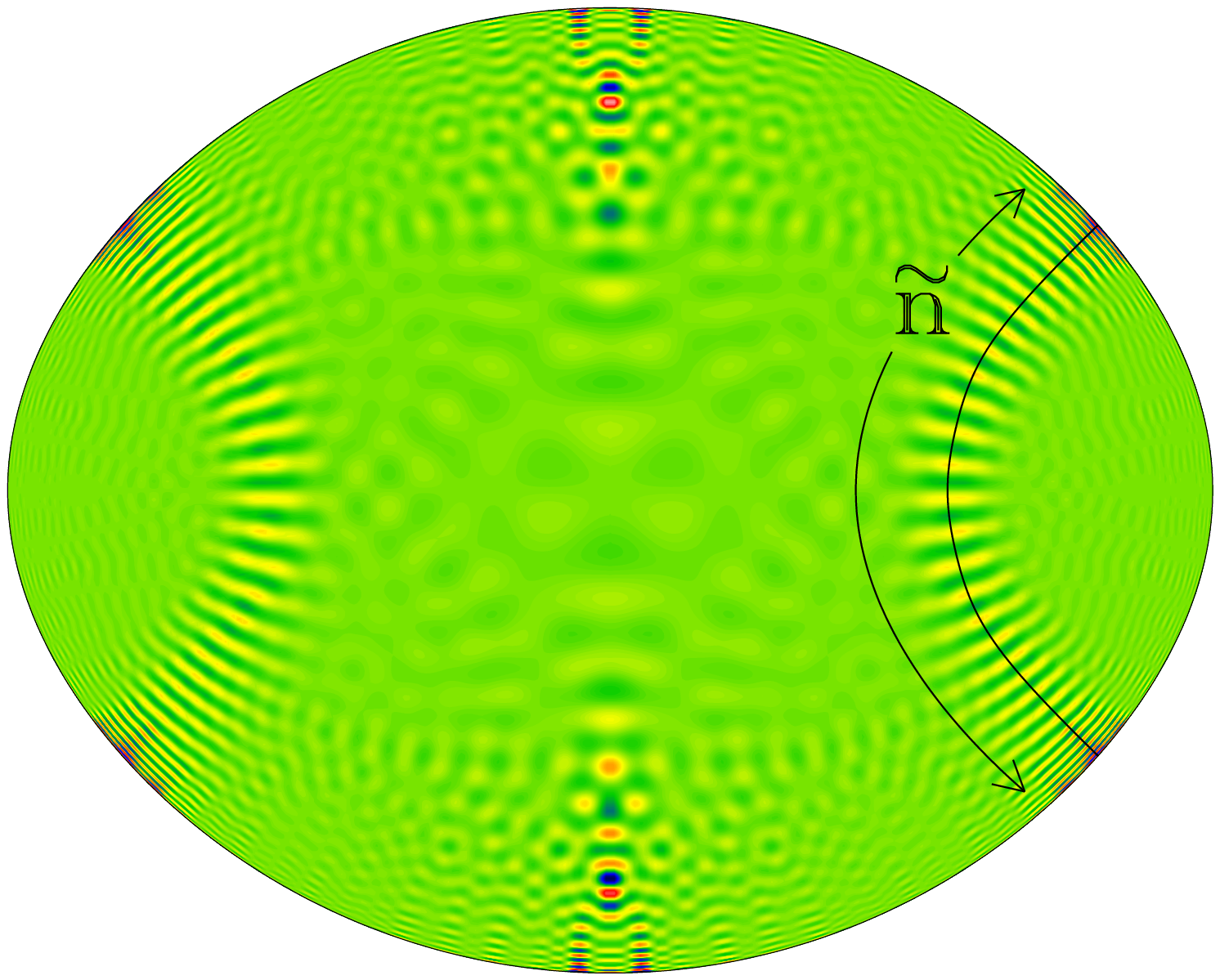} $\quad$
\includegraphics[height=4cm]{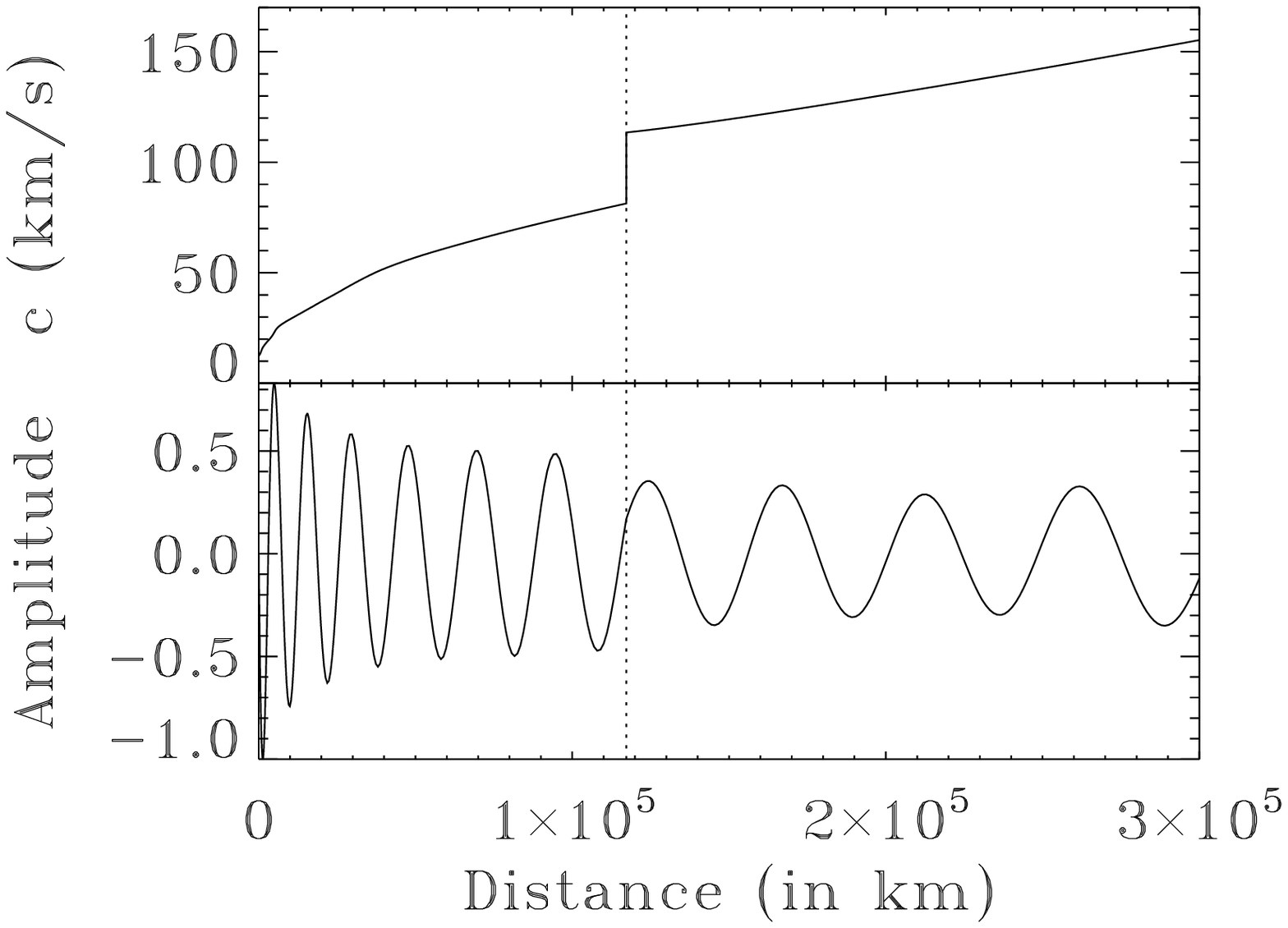}
\end{center}
\caption{\textit{Left:} meridional cross-section of an island mode.
\textit{Right:} sound velocity and mode profile along the path shown in the left
panel (only part of the profile is shown for legibility).
\label{fig:mode_profile}}
\end{figure}

We first turn our attention to the effects of discontinuities on the
eigenfunctions.  Figure~\ref{fig:mode_profile} shows the meridional
cross-section of an island mode as well as the sound velocity and mode profile
along a heuristically determined path.  As can be seen in the right panel, the
discontinuity modifies the wavelength as well as the amplitude of the
oscillations.  Further tests confirm that the wavelength scales with the sound
velocity.  Another effect which has already been pointed out in
\citet{Reese2011} is a slight deviation of the mode at the discontinuity.

\begin{figure}[htbp]
\begin{center}
\includegraphics[width=0.8\textwidth]{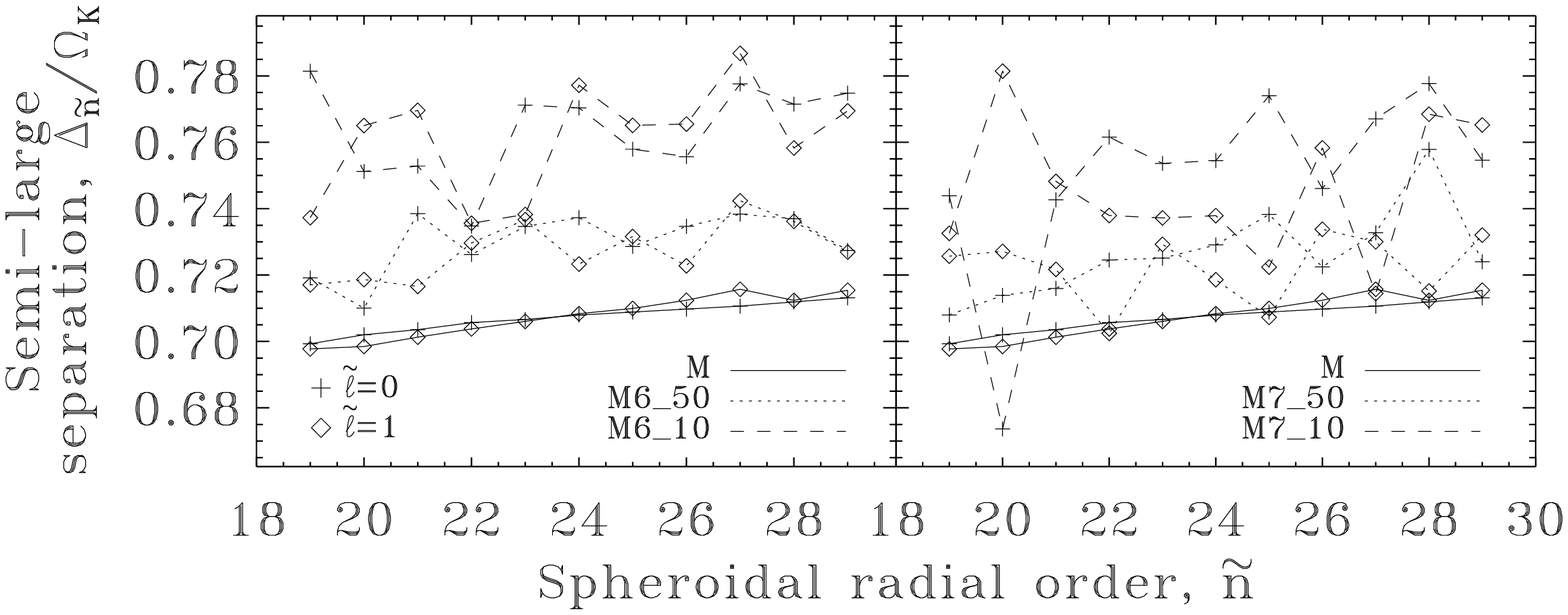}
\end{center}
\caption{Semi-large frequency separation, $\Delta_{\tilde{n}} =
\nu_{\tilde{n}+1} - \nu_{\tilde{n}}$, for axisymmetric $(m=0)$ modes, as a
function of $\tilde{n}$, the spheroidal radial order (see
Fig.~\ref{fig:mode_profile}, left panel). \label{fig:delta_n}}
\end{figure}

At low rotation rates, discontinuities affect the frequencies by superimposing
an oscillatory pattern over the usual frequency spectrum \citep[\textit{e.g.}
][]{Monteiro1994}.  A similar effect takes place here, as illustrated by the
semi-large frequency separations shown in Fig.~\ref{fig:delta_n}, although the
oscillatory pattern is less regular.  One can also calculate the scatter between
the numerical frequencies and a simplified version of the asymptotic formula
\citep[see Eq. (27) of][]{Reese2009a}.  The scatter,
$\left<(\nu_{\mathrm{asymp.}}-\nu)^2\right>^{1/2}/\Delta_{\tilde{n}}$, ranges
from 0.0143 for model \texttt{M} to 0.0436 for model \texttt{M7\_10}.  Even in
the best case, the scatter is more than an order of magnitude larger than the
scatter obtained around the main sequence of equidistant frequencies found in
\citet{Breger2012}, thereby supporting the conclusion that this sequence is not
caused by an asymptotic behaviour.


\begin{figure}[htbp]
\begin{center}
\includegraphics[width=0.8\textwidth]{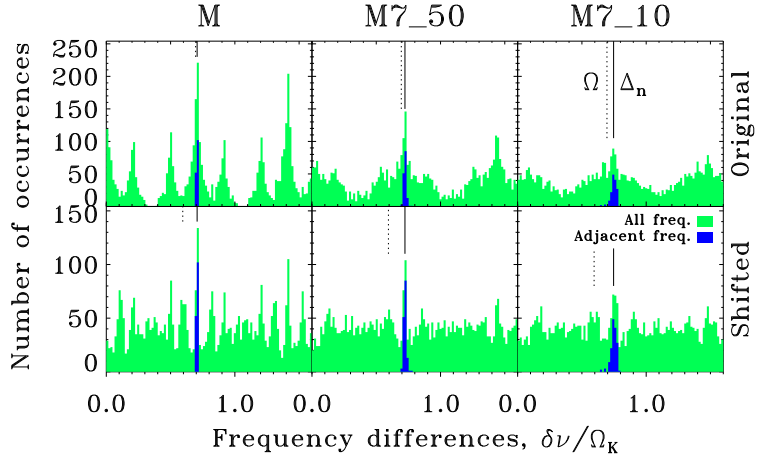}
\end{center}
\caption{Histograms of frequency differences for three models (see text for
details).\label{fig:histograms}}
\end{figure}

One can then investigate whether it is possible to recover the semi-large
frequency separation, $\Delta_{\tilde{n}}$.  Figure~\ref{fig:histograms} shows
histograms of frequencies differences for three models.  The lightly shaded
areas show all frequency differences, whereas the dark areas show the frequency
differences from modes with adjacent $\tilde{n}$ values and the same
$(\tilde{\ell},m)$ values.  The upper row is based on the original numerical
frequencies.  In all cases, the semi-large frequency, $\Delta_{\tilde{n}}$
separation shows up clearly.  However, it turns out that the rotation rate is
close to $\Delta_{\tilde{n}}$ thereby amplifying the signal, due to island mode
multiplets \citep{Pasek2012}.  This can be seen by comparing the light and dark
regions in the histograms.  In the lower row, the frequencies were shifted by
$0.1m$, thereby mimicking a lower rotation rate.  Even in this situation, a peak
remains at $\Delta_{\tilde{n}}$ for all three models.

In conclusion, although discontinuities lead to more scatter around the
asymptotic behaviour of island modes and may complicate mode identification,
they are unable to mask features such as the semi-large frequency separation. 
Hence, the asymptotic formula remains a viable explanation for stars such as
those observed by \citet{GarciaHernandez2009, GarciaHernandez2013}.

\section*{Acknowledgements}
DRR is financially supported through a postdoctoral fellowship from the
``Subside fédéral pour la recherche 2012'', University of Liège. FEL and MR
acknowledge the support of the French Agence Nationale de
la Recherche (ANR), under grant ESTER (ANR-09-BLAN-0140).

\end{document}